\documentclass[fleqn,usenatbib]{mnras}
\usepackage{newtxtext,newtxmath}
\usepackage[T1]{fontenc}

\DeclareRobustCommand{\VAN}[3]{#2}
\let\VANthebibliography\thebibliography
\def\thebibliography{\DeclareRobustCommand{\VAN}[3]{##3}\VANthebibliography}

\usepackage{graphicx}	% Including figure files
\usepackage{amsmath}	% Advanced maths commands
\usepackage{multicol}        % Multi-column entries in tables
\usepackage{bm}		% Bold maths symbols, including upright Greek
\usepackage{pdflscape}	% Landscape pages
\usepackage{longtable}  % for long tables
\usepackage{caption} % caption for non-floating figure
\usepackage{subfigure} % for figure grids of single caption
\usepackage[flushleft]{threeparttable}
\usepackage{multirow}   % for multicolumn tables
\usepackage{xfrac}   % for \sfrac macro
\usepackage{xcolor} % for text colours
\usepackage{todonotes} % for \todo notes
\usepackage{outlines} %nested lists
\usepackage{ulem} % strikeout
\usepackage{siunitx}

\newcommand{\ignore}[1]{}
\newcommand{\dm}{~pc~cm$^{-3}$} %               dm units
\newcommand\xmmsrc{2XMM J104608.7$-$594306}
\title[Bright radio burst from 2XMM~J1046]{A bright wideband radio burst from the isolated neutron star 2XMM J104608.7$-$594306}

\author[J. Tian et al.]{J. Tian$^{1}$\thanks{E-mail: jun.tian@manchester.ac.uk}, K. M. Rajwade$^2$\thanks{E-mail: kaustubh.rajwade@physics.ox.ac.uk}, I. Pastor-Marazuela$^{5,6,1}$, B. W. Stappers$^1$, M. Caleb$^{3,4}$, K. Shaji$^{3,4}$, S. Singh$^{1}$,
\newauthor E. D. Barr$^{7}$, M. Kramer$^{7}$\\
$^{1}$ Jodrell Bank Centre for Astrophysics, Department of Physics and Astronomy, The University of Manchester, Manchester M13 9PL, UK \\
$^2$Astrophysics, The University of Oxford, Denys Wilkinson Building, Keble Road,
Oxford OX1 3RH, UK \\
$^3$Sydney Institute for Astronomy, School of Physics, The University of Sydney, NSW 2006, Australia \\
$^4$ARC Centre of Excellence for Gravitational Wave Discovery (OzGrav), Hawthorn, 3122, Victoria, Australia \\
$^{5}$ ASTRON, the Netherlands Institute for Radio Astronomy, Oude Hoogeveensedijk 4,7991 PD Dwingeloo, The Netherlands\\
$^{6}$ Anton Pannekoek Institute, University of Amsterdam, Amsterdam, The Netherlands\\
$^{7}$Max-Planck-Institut f{\"u}r Radioastronomie, 53121 Bonn, Germany
}

\date{Accepted XXX. Received YYY; in original form ZZZ}

\pubyear{2025}

\begin{document}
\label{firstpage}
\pagerange{\pageref{firstpage}--\pageref{lastpage}}
\maketitle

% Abstract of the paper
\begin{abstract}
We present the discovery of a second coherent radio burst from the thermally emitting neutron star \xmmsrc~in our follow-up observations with the Murriyang Ultra-Wideband Low receiver. This burst shows complex morphology with multiple components and wideband emission spanning from 704 to 4032\,MHz. We measured a steep spectral index of $\alpha=-2.18\pm0.16$. Our polarimetric analysis demonstrates that the burst is highly polarised with a linear and circular polarisation fraction of 54\% and 22\%, respectively. We identified an orthogonal jump in the polarisation position angles of the burst, resembling those seen in radio pulsars. We compared this burst with the first radio burst detected from the source with MeerKAT. These two bursts detected in a total of 40 hours on source with MeerKAT and Murriyang, combined, show that \xmmsrc\ can emit sporadic radio emission with luminosity jumps comparable to those seen in the bright bursts from SGR 1935+2154.
%The detection of a second burst shows that \xmmsrc\ can emit sporadic radio emission with a luminosity comparable to some of the bright bursts seen from SGR 1935+2154, 
This suggests that previously thought radio-quiet neutron stars such as X-ray dim isolated neutron stars and central compact objects could exhibit rare radio bursting activity.
\end{abstract}

% Select between one and six entries from the list of approved keywords.
% Don't make up new ones.

%%%%%%%%%%%%%%%%%%%%%%%%%%%%%%%%%%%%%%%%%%%%%%%%%%

%%%%%%%%%%%%%%%%% BODY OF PAPER %%%%%%%%%%%%%%%%%%

\section{Introduction}\label{intro}

The neutron star (NS) zoo consists of different types of NSs, including rotation-powered radio pulsars, highly magnetised NSs (magnetars), X-ray dim isolated NSs (XDINSs) and central compact objects (CCOs). As individual classes, they differ in observational characteristics such as their electromagnetic spectrum and time intermittency and variability. While radio emission is prevalent in the manifestation of NSs, as observed from radio pulsars and some magnetars, until recently XDINSs and CCOs were thought to be radio-quiet %despite extensive searches for their radio emission 
\citep{Kondratiev09, Zane11, Pastor23, Lu24, Kurpas25}. 
The discovery of very faint radio emission from the CCO 1E 1207.4$-$5209 in the supernova remnant G296.5+10.0 challenged this paradigm \citep{Zhang25}.
In contrast, the recent serendipitous discovery of a coherent radio burst with MeerKAT from 2XMM J104608.7$-$594306, an XDINS candidate discovered by XMM-Newton through its thermal X-ray emission \citep{Pires09, Pires12, Pires15}, shows that XDINSs are capable of producing sporadic radio bursts similar to those seen from magnetars \citep{Rajwade25}.

XDINSs are characterised by quasi-blackbody soft X-ray emission without persistent radio counterparts and lack of association to a supernova remnant. The first seven XDINSs, known as ‘The Magnificent Seven’, were discovered in ROSAT All-Sky Survey (RASS) data and located at a distance of $\lesssim500$\,pc \citep{Haberl07}. They share similar properties, including thermal X-ray spectra with effective temperatures of $\sim40\text{--}100$\,eV, low X-ray luminosities of $\sim10^{31}\,\text{erg}\,\text{s}^{-1}$, low absorption column densities of $\sim10^{20}\,\text{cm}^{-2}$, long rotation periods of $\sim3\text{--}10$\,s, strong magnetic fields of $\sim10^{13}$\,G and very faint optical counterparts with $B$-band magnitudes $\gtrsim25$ (corresponding to X-ray-to-optical flux ratios $\gtrsim10^4$). 
\xmmsrc\ is an X-ray source discovered by XMM-Newton in the Carina Nebula \citep{Pires09}. The X-ray observations revealed a purely thermal spectrum with a blackbody temperature of $\approx135$\,eV and an absorbing column of $\approx3\times10^{21}\,\text{cm}^{-2}$. Deep optical follow-up observations yielded a very high X-ray-to-optical flux ratio of $>10^{3.8}$ \citep{Pires12}. These properties are remarkably reminiscent of XDINSs though no X-ray pulsation was firmly detected \citep{Pires15}.
The first confirmed radio emission from 2XMM J104608.7$-$594306 \citep{Rajwade25} suggests that the dearth of radio detection of XDINSs could be caused by %their radiation beam pointing away from us rather than the radio emission being suppressed. 
unfavorable beaming geometry or highly intermittent emission, rather than intrinsic suppression of the radio emission.
This is also supported by the similar X-ray properties observed between a persistent radio pulsar PSR J0726$-$2612 and XDINSs \citep{Rigoselli19}.

XDINSs could be evolutionarily associated with magnetars given their similarly high X-ray luminosities and long spin periods \citep{Kaspi10, Yoneyama19}. The analysis of the bright radio burst detected from 2XMM J104608.7$-$594306 revealed more similarities between this source and the magnetar population \citep{Rajwade25}. Its radio luminosity reached $\sim4.6\times10^{29}\,\text{erg}\,\text{s}^{-1}$, similar to the average radio luminosity of radio-emitting magnetars \citep{Esposito21}. The temporal structure observed in this burst consists of at least eight distinct sub-components, resembling the quasi-periodic microstructure seen in magnetars and other NS populations \citep{Kramer24}. Finding more radio bursts from 2XMM J104608.7$-$594306 could potentially reveal more similarities between this object and magnetars. In addition, since magnetars are thought to be plausible progenitors of fast radio bursts (FRBs; \citealt{Bochenek20, CHIME20}), XDINSs might be able to produce FRB-like bursts. However, we need to find radio bursts much brighter than the one detected from 2XMM J104608.7$-$594306 to test this scenario.

In this work, we present further follow-up observations on studies of 2XMM J104608.7$-$594306 with the ultrawide bandwidth low-frequency (UWL; 704--4032\,MHz) receiver on the 64-m Murriyang radio telescope \citep{Hobbs20}. We detect a second radio burst from 2XMM J104608.7$-$594306, which is brighter than the first one and shows extremely broadband emission that spans the full bandwidth of the UWL. In Section~\ref{sec:obs}, we describe the Murriyang observations and data reduction. Our results are then presented in Section~\ref{sec:results}. We discuss the implications of our results in Section~\ref{sec:disc}, followed by a summary in Section~\ref{sec:sum}.

\section{Observations and Data Reduction}\label{sec:obs}

\begin{figure*}
	\centering
	\includegraphics[width=0.7\textwidth]{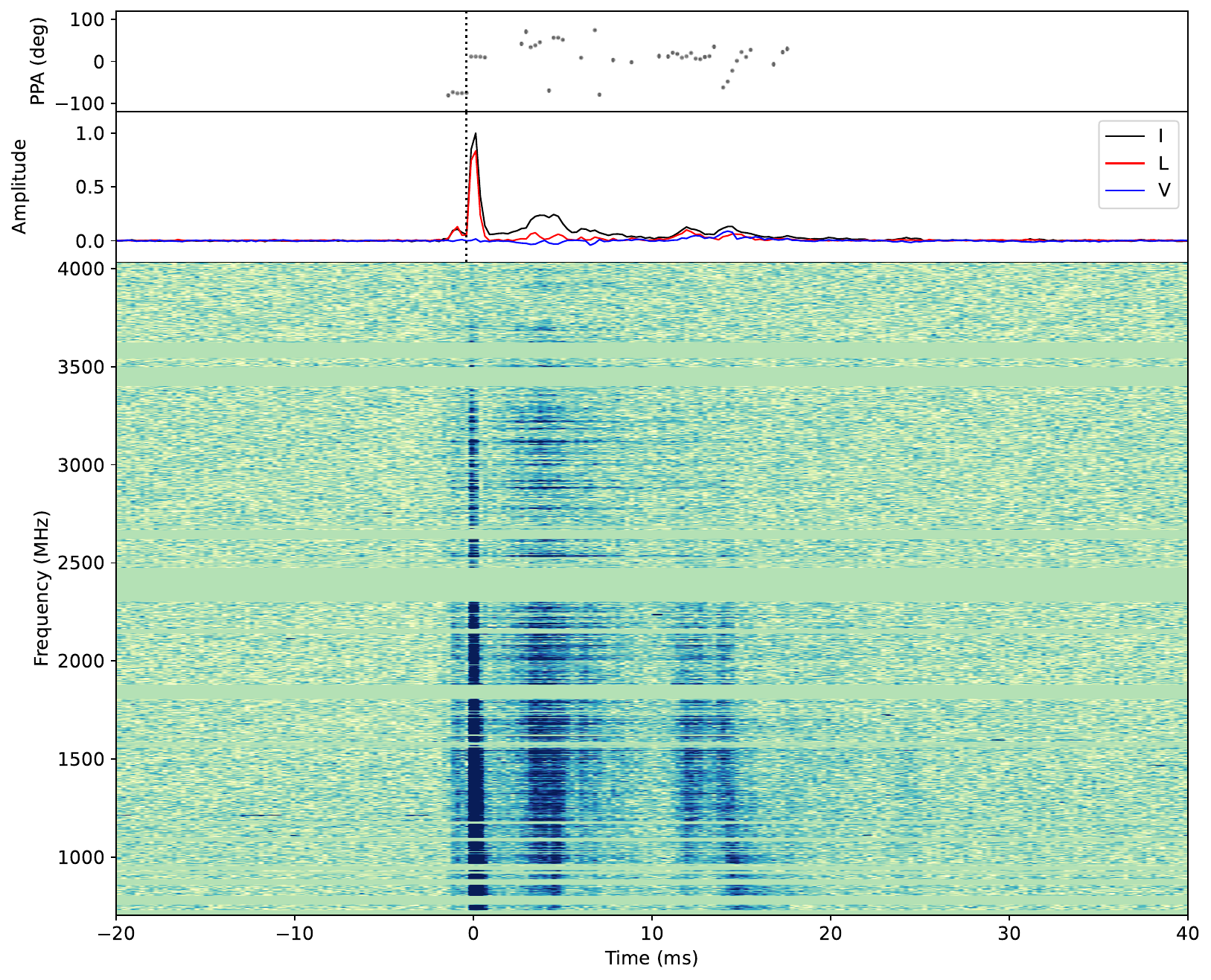}
	\caption{Murriyang/Parkes detected radio burst from \xmmsrc. The bottom panel shows the dynamic spectrum, the middle panel shows the frequency averaged pulse profile for total intensity ($I$), linear polarisation ($L$) and circular polarisation ($V$), and the top panel shows the polarisation position angle. The pulse has been coherently dedispersed to the structure maximising DM of $99.40\,\text{pc}\,\text{cm}^{-3}$, and the polarisation data have been Faraday corrected to the rotation measure of $-49.3\,\text{rad}\,\text{m}^{-2}$. The dotted vertical line in the top and middle panel indicates the time of the orthogonal jump. Blank horizontal lines in the bottom panel indicate data which were flagged due to the presence of RFI.
	}
	\label{fig:pulse}
\end{figure*}

We followed up 2XMM J104608.7$-$594306 with the Murriyang/Parkes UWL for a total of 36 hours between April 2025 and November 2025. The date and duration of each observation are shown in Figure~\ref{fig:obs}. %We also include previous MeerKAT observations of this field recorded by the MeerTRAP instrument, a commensal fast radio transient detection programme started in 2019 \citep{Sanidas18, Bezuidenhout22, Rajwade22, Caleb23, Jankowski23, Driessen24}.
As can be seen, most observations lasted 1--2\,hr on a single day, and the longest observation took place on 8 October 2025 with a duration of 6\,hr.
The Murriyang observations were recorded in the pulsar search mode with a sampling time of $256\,\mu$s and a frequency resolution of 1\,MHz (i.e. 3328 channels across the full bandwidth). The data were coherently dedispersed to the dispersion measure (DM) of $98.41\,\text{pc}\,\text{cm}^{-3}$ and stored in 8-bit PSRFITS files \citep{Hotan04} with full Stokes information. A 2-min scan of a linearly polarised noise diode was performed prior to each observation.

%\begin{figure*}
%\centering
%\includegraphics[width=0.7\textwidth]{MTP0096_Parkes_obs.pdf}
%\caption{A timeline of targeted Parkes observations of 2XMM J104608.7$-$594306. Each vertical line shows the time of an observation, and the height of the line shows the observation duration. The burst detection is marked by a red star.}
%\label{fig:obs}
%\end{figure*}

We split the full band of the Murriyang observations evenly into eight sub-bands, each with a bandwidth of 416\,MHz.
We searched the sub-banded data for single pulses with $\pm10\,\text{pc}\,\text{cm}^{-3}$ around the nominal DM of $98.41\,\text{pc}\,\text{cm}^{-3}$ using {\sc transientx}\footnote{\url{https://github.com/ypmen/TransientX}} \citep{TransientX}. The DM step size was set to $0.1\,\text{pc}\,\text{cm}^{-3}$, and the maximum boxcar width to 100\,ms. We visually inspected all candidates with a peak $\text{S/N}>8$ (as defined in~\citealt{lorimer2004}) and identified a single bright pulse on 7 April 2025, as indicated in Figure~\ref{fig:obs}. In order to perform a cross-check, we repeated the single pulse search using {\sc presto} \citep{Ransom01}. We redetected the bright pulse on 7 April 2025, and no other pulses were found. We checked the notices of all-sky high-energy monitors including \textit{Swift}-BAT \citep{Krimm13}, \textit{Fermi}-GBM \citep{Meegan09} and INTEGRAL \citep{Jensen03}, and did not find high-energy flares at the time of the Murriyang burst.

We searched each of the Murriyang observations with the Fast Folding Algorithm (FFA; \citealt{FFA_morello}) as \xmmsrc\ may emit weak regular periodic signals. The data were cleaned and dedispersed to the nominal DM using {\sc transientx}, and then searched using the {\sc rffa} utility of the {\sc riptide}\footnote{\url{https://github.com/v- morello/riptide}} package over a period range of 0.1--100\,s and a duty-cycle range of 0.2\%--20\%. We folded all candidates with $\text{S/N}>8$ for manual inspection, and found no periodic signals of astrophysical origin in any of the Murriyang observations. We also conducted a periodicity search using the {\sc accelsearch} routine from {\sc presto}\footnote{\url{https://github.com/scottransom/presto}} \citep{Ransom01}, summing up to 32 harmonics. The resulting candidates were folded and visually inspected, and no astrophysical signals were found. However, using the longest Murriyang observation allowed us to place deep constraints on any periodic emission from \xmmsrc. Given that the system equivalent flux density of the UWL receiver on Murriyang was measured to be 33--72\,Jy depending on the frequency \citep{Hobbs20}, the 6\,hr observation on 8 October 2025 corresponds to a sensitivity limit of $\sim50\mu$Jy for an $8\sigma$ detection threshold based on the radiometer equation, considering a duty cycle of 10\% and an effective bandwidth of 773\,MHz (excluding flagged channels) within the lower UWL band (0.7 to 1.7\,GHz).

\section{Results}\label{sec:results}

We discovered a bright radio burst in the 36\,hr follow-up observations of \xmmsrc\ with Murriyang, as shown in Figure~\ref{fig:pulse}. Considering the complex morphology and sub-components in the burst, we used {\sc dm\_phase}\footnote{\url{https://github.com/danielemichilli/DM_phase}} \citep{Seymour19}, a DM optimisation algorithm that maximises the coherent power across the bandwidth, to measure the DM. We dedispersed the burst over a trial DM range of 95--105\dm with a step size of 0.1\dm, and obtained a structure-maximising DM of $99.40\pm0.07$\dm. This value is consistent with the DM measurement for the MeerKAT burst within $1\sigma$ \citep{Rajwade25}.

% \begin{figure*}
% 	\centering
% 	\includegraphics[width=0.7\textwidth]{MTP0096_pulse.pdf}
% 	\caption{Murriyang/Parkes detected radio burst from \xmmsrc. The bottom panel shows the dynamic spectrum, the middle panel shows the frequency averaged pulse profile for total intensity ($I$), linear polarisation ($L$) and circular polarisation ($V$), and the top panel shows the polarisation position angle. The pulse has been coherently dedispersed to the structure maximising DM of $99.40\,\text{pc}\,\text{cm}^{-3}$, and the polarisation data have been Faraday corrected to the rotation measure of $-48.7\,\text{rad}\,\text{m}^{-2}$. The dotted vertical line in the top and middle panel indicates the time of the orthogonal jump. Blank horizontal lines in the bottom panel are flagged due to RFI.
% 	}
% 	\label{fig:pulse}
% \end{figure*}

\subsection{Burst morphology}

The Murriyang burst from \xmmsrc\ shows a similar unusual emission profile to the MeerKAT burst \citep{Rajwade25}: an extremely sharp rise and a slowly decaying morphology, as shown in Figure~\ref{fig:pulse}. Yet, the Murriyang burst comprises more sub-components, including a precursor prior to the brightest component, which is not observed in the MeerKAT burst (see Figure~\ref{fig:compare} for a comparison). This precursor is about nine times fainter than the brightest component. Such a feature would be invisible in the MeerKAT data due to the lower S/N. Both bursts from \xmmsrc\ show temporal structures similar to the quasi-periodic microstructure seen in radio-emitting neutron stars \citep{pastor-marazuela_fast_2023, Kramer24}, which may reflect disturbances in the magnetosphere induced by activity in the neutron star crust \citep{Wadiasingh20} or intrinsic angular patterns in the beamed radio emission \citep{Kramer02}. 

% IPM
To investigate the microstructure observed in the Murriyang burst, we generated the power spectrum of the burst pulse profile using \texttt{Stingray} \citep{bachetti_stingray_2024}. We modeled the spectrum with both a single and a broken power law, each including a white noise component.%, as shown in Fig.~\ref{fig:powspec}. 
The broken power law model is preferred to fit the power spectrum, resulting in lower Bayesian (BIC) and Akaike (AIC) information criteria. In contrast, the single power law model requires a negative white noise constant, which is non-physical and indicates that the curvature of the power spectrum cannot be adequately captured with a simple power law. 
The break frequency in the broken power law model is located at $0.9\pm0.1$\,kHz, corresponding to a characteristic timescale of $1.1\pm0.1$\,ms. This suggests the pulse profile has a typical microstructure timescale rather than a strict periodicity.

We also performed a periodicity search by fitting the times of arrival (ToAs) of the burst components as a function of component number with a linear function, allowing for gaps between components, following the method described in \citet{pastor-marazuela_fast_2023}. Because of the complexity of the burst morphology, the ToAs were taken to be the peak times of the pulse profile components, with the time resolution adopted as the uncertainty, rather than fitting a multi-component Gaussian model. This procedure identified 12 components with gaps, corresponding to the component index vector $n=(0, 1, 2, 3, 4, 5, 7, 8, 10, 11, 13, 17)$ and a mean separation of 1.91\,ms. To assess the significance of the periodicity, we simulated 10000 sets of random component ToAs with waiting times drawn from a Poissonian distribution with the same mean spacing and an exclusion parameter $\eta=0.2$. We fitted those simulated ToAs to the same linear function as the data, and compared the reduced $\chi^2$ statistic. The resulting significance is $2.2\sigma$, well below the $3\sigma$ threshold typically required to claim a periodic signal.

Since the characteristic timescale inferred from the power spectrum analysis differs from the inferred mean component separation obtained from the ToA fitting, and the latter periodicity test results in a significance below the $3\sigma$ threshold, the burst structure does not show evidence for a strictly periodic structure.

% We also ran a periodicity search by fitting multiple components to the pulse profile. This resulted in 12 components with a mean separation of 1.91\,ms, and we found no periodicity above $3\sigma$ between these components.

%\begin{figure}
%    \centering
%    \includegraphics[width=\linewidth]{MTP0096_pl_fit.pdf}
%    \caption{Power spectrum of the pulse profile. The top panel shows the power spectrum (black), fitted to a power law (blue dashed line) and a broken power law (red solid line) function, with the break frequency shown as a grey dashed line with a shaded area. The power spectrum of the noise is shown in grey as a comparison. The bottom panel shows the fit residuals for each function.}
%    \label{fig:powspec}
%\end{figure}

We do not see evidence of scattering in the brightest and narrowest component of the Murriyang burst (see Figure~\ref{fig:pulse}), consistent with that observed in the MeerKAT burst. Any underlying scattering in the radio emission should have a timescale shorter than the width of the narrowest component (0.5\,ms).
 
% IPM - feel free to rephrase or move to discussion
% We notice that the last component visible on the dynamic spectrum at around 15\,ms in Fig.~\ref{fig:pulse} displays a downward drifting slope. This, in addition to the lower frequencies at which the two latter components are visible, resembles the morphology that is typically observed in repeating FRBs. While the bandwidth of the Parkes detection presented here is much larger than that of typical blind FRB surveys, these similarities may suggest that the emission processes responsible for this burst structure are related to those operating in known extragalactic repeating FRBs, potentially pointing to a common emission mechanism or evolutionary path.

\subsection{Spectrum}\label{sec:spec}

To measure the flux density of the burst, we calibrated the Murriyang data using the flux calibrator PKS B1934$-$638, following the method described in \citet{Dai19}. We measured the peak flux density to be $30.03\pm0.12$\,Jy, where the uncertainty was estimated using the standard deviation of the baseline noise. This is $\sim100\times$ brighter than the MeerKAT burst, and corresponds to a radio luminosity of $\sim2\times10^{32}\,\text{erg}\,\text{s}^{-1}$ assuming isotropic emission and a source distance of 1.5\,kpc. %$\sim100$ times brighter than the MeerKAT detected pulse. 
We defined the on-pulse region as the interval between the first and last sample with a flux density greater than three times the noise level, and obtained a burst duration of 33.5\,ms. Integrating the flux density over the on-pulse region we obtained a fluence of $66.4\pm4.0$\,Jy\,ms.

We analysed the spectrum of the entire burst by dividing the Murriyang data into $13\times256$\,MHz sub-bands and measuring the flux density in each sub-band. We assumed that the burst duration does not change across the sub-bands. We also assumed that each sub-band contains many scintles that our analysis is not limited by scintillation. This is likely to be true given that the NE2001 model \citep{ne2001} predicts a scintillation bandwidth of only $\sim200$\,Hz at 1\,GHz along the line of sight of \xmmsrc. %The averaged flux density over the on-pulse region at each subband is shown in Figure~\ref{fig:spectrum}. Due to the faint emission in the two highest subbands, we provide only $3\sigma$ upper limits for these frequencies. 
We fitted the spectrum with a simple power-law function, $S_\nu\propto\nu^\alpha$, and obtained a spectral index of $\alpha=-2.18\pm0.16$. %, as shown in Figure~\ref{fig:spectrum}. 
We also measured the spectral indices of the different components in the burst and obtained values ranging between $-2.63\pm0.16$ and $-1.72\pm0.11$. No monotonic evolution of the spectral index was observed. %with no evidence of spectral evolution.
%We also measured the spectral index of the brightest component in the burst, and obtained a value of $-2.50\pm0.03$.

%\begin{figure}
%	\centering
%	\includegraphics[width=0.5\textwidth]{spectral_index.png}
%	\caption{Averaged flux density over the on-pulse region as a function of frequency. The red line shows the best-fitting power law for the spectrum.
%	}
%	\label{fig:spectrum}
%\end{figure}

\subsection{Polarimetry}

We created a full-Stokes {\sc Psrfits} format archive for the burst by extracting the Murriyang data using {\sc dspsr} \citep{Straten11}. Approximately 25\% of the 3328 frequency channels were flagged using {\sc Psrchive} \citep{Hotan04} due to radio frequency interference (RFI). The data were polarisation calibrated using the 2-min noise diode observation. We measured the Faraday rotation in the Stokes data using the {\sc RMsynth} tool from the {\sc Psrsalsa}\footnote{\url{https://github.com/weltevrede/psrsalsa}} software suite \citep{Weltevrede16}. We searched a broad range of rotation measures (RMs) between $\pm10000\,\text{rad}\,\text{m}^{-2}$ with a step size of $0.1\,\text{rad}\,\text{m}^{-2}$, and obtained a value of $-49.3\pm0.2\,\text{rad}\,\text{m}^{-2}$. This is consistent with the RM measurement for the MeerKAT burst within $1\sigma$ \citep{Rajwade25}.

After correcting for the Faraday rotation, we created the polarimetric pulse profile using {\sc psrsalsa}, as shown in Figure~\ref{fig:pulse}. Bias in the linear polarisation $L$ was removed for each sample following the method in \citet{Wardle74}, and the polarisation position angles (PPAs) were measured for samples with $L>3\sigma$, where $\sigma$ is the off-pulse noise. We averaged $L/I$ and $|V|/I$ across the pulse profile and obtained a linear and circular polarisation fraction of $0.54\pm0.02$ and $0.22\pm0.01$, respectively. Parts of the burst show nearly 100\% polarisation (e.g., in the precursor component), suggesting that the emission is produced by coherent curvature radiation \citep{Rahaman22}. We see variations in the PPA across the pulse profile in Figure~\ref{fig:pulse}. In particular, there is a PPA jump between the precursor and the brightest component, as indicated by the dotted vertical line. We used the mean values of PPAs before and after the jump to calculate the jump angle. This resulted in a value of $\ang{87.2}\pm\ang{1.4}$. The uncertainty was computed from the standard deviation of PPAs weighted by the errors in the PPA measurements and based on the principle of error propagation. The observed PPA jump is consistent with an orthogonal jump within the $2\sigma$ error range, which has been observed in many pulsars \citep{Manchester75, Cordes78, Backer80, Stinebring84, Karastergiou09} and some FRBs \citep{Niu24, Jiang24}. This further corroborates the neutron star nature of \xmmsrc.

\section{Discussion}\label{sec:disc}

\xmmsrc\ exhibits many characteristics similar to those seen in radio-loud magnetars: transient radio emission, complex burst morphology with multiple emission components, high linear polarisation, and significant brightness variability between the Murriyang and MeerKAT bursts. The spectral index we measured is steeper than the spectral indices of most magnetars that typically range from $-0.5$ to $+0.3$ \citep{Lazaridis08, Torne15, Dai19}, but similar to Swift J1818.0$-$1607 \citep{Lower20}. Meanwhile, the radio luminosities of the Murriyang and MeerKAT bursts are comparable to the average radio luminosity of radio-loud magnetars \citep{magnetar_review_kaspi, Esposito21}. %, and the X-ray luminosity of \xmmsrc\ is consistent with those of XDINSs. 
These suggest that \xmmsrc\ could represent an evolutionary link between XDINSs and magnetars. 

\begin{figure*}
	\centering
	\includegraphics[width=0.6\textwidth]{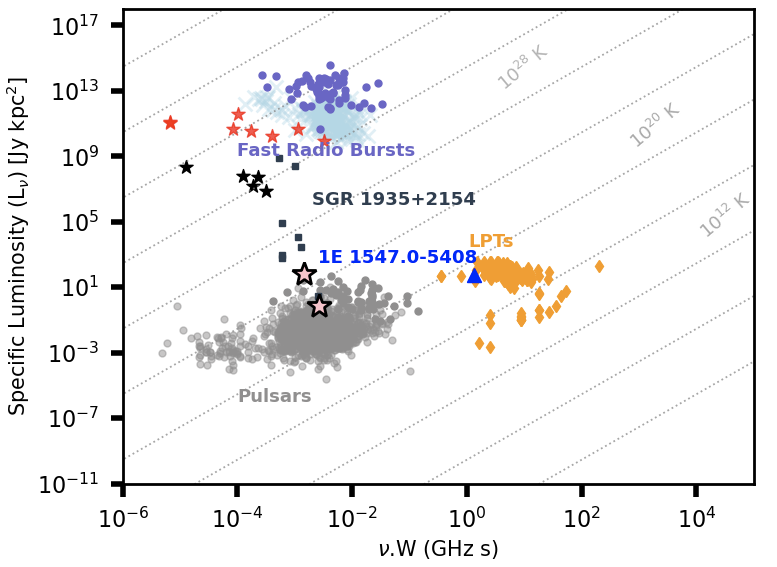}
	\caption{Phase space of various coherent radio transients. The brightest components of the two bursts detected by MeerKAT and Murriyang from \xmmsrc\ are shown by the pink stars. The dotted grey lines represent lines of constant brightness temperatures. The data points for FRB 20180916B (red stars), FRB 20121102A (light blue crosses), FRB 20200112E (black stars), one-off FRBs (purple points), SGR 1935+2154 (black squares), 1E 1547.0$-$5408 (blue triangle), pulsars (gray points) and LPTs (yellow diamonds) are taken from \citet{Caleb24}, \citet{israel2021}, \citet{Nimmo22} and references therein.
	}
	\label{fig:phase_space}
\end{figure*}

Alternatively, considering the X-ray luminosity of \xmmsrc\ is consistent with that of CCOs \citep{Luca17}, \xmmsrc\ could be a CCO candidate. However, identifying a supernova remnant in the Carina Nebula is difficult due to interactions of strong winds of massive stars with their environments and the extinction at optical and infrared wavelengths \citep{Townsley11}. Given that most CCOs show fast spin periods of $0.1\text{--}0.4$\,s and weak dipolar magnetic fields of $\sim10^{11}$\,G \citep{Zavlin00, Gotthelf05, Gotthelf09}, finding the spin period and period derivative of \xmmsrc\ will allow us to confirm its nature. We did not find persistent pulsed radio emission above $\sim50\,\mu$Jy in the Murriyang observations. This limit is comparable to the flux density of the faint pulsed radio emission recently detected from the CCO 1E 1207.4$-$5209 ($33\,\mu$Jy at 816\,MHz; \citealt{Zhang25}). Also, \xmmsrc\ and 1E 1207.4$-$5209 are located at similar distances. Given these considerations, we cannot rule out the presence of radio pulsations from \xmmsrc, and more sensitive observations in the time domain are required to probe this further.

We identified an orthogonal jump in the PPA of the Murriyang burst. %This may be explained by the superposition of the O-mode and X-mode waves that are separated by $90^\circ$ in the highly magnetised magnetosphere of a pulsar \citep{Barnard86}. 
This is similar to the orthogonal PPA jumps seen in pulsars, which can be explained by the superposition of the O-mode and X-mode waves that are separated by $90^\circ$ in the highly magnetised magnetosphere \citep{Barnard86}.
Both coherent and incoherent superpositions of the two orthogonal modes are possible. While the coherent superposition allows for non-orthogonal jumps and the existence of circular polarisation \citep{Dyks17}, the incoherent superposition predicts depolarisation at the jump time \citep{Singh24}. We do not observe significant depolarisation when the PPA jump happens in Figure~\ref{fig:pulse}. This may be attributed to our limited sampling time of $256\,\mu$s, which is insufficient to resolve the PPA jump. Consequently, we cannot distinguish between the coherent and incoherent superpositions for the PPA jump observed here. %However, we cannot rule out the possibility of  The latter matches the observation in Figure~\ref{fig:pulse}, i.e. depolarisation and no circular polarisation when the PPA jump happens.
We use the short timescale of the PPA jump to derive a generic constraint on the emission region. The timescale of $<256\,\mu$s corresponds to a light-crossing length of $<77$\,km. This is much smaller than the light cylinder radius of typical pulsars \citep{lorimer2004}, consistent with the radio emission arising from the neutron star magnetosphere.

The dynamic spectrum of the second radio burst reveals peculiar radio emission that is reminiscent of some features that are exhibited by repeating FRBs. For example, similar to the discovery burst, the dynamic spectrum shows multiple emission components with subsequent emission components becoming visible at lower radio frequencies. In the discovery paper,~\cite{Rajwade25} discuss this behaviour as 
a manifestation of localised emission components having different spectral indices. However, since there is no evidence of spectral evolution here (see Section~\ref{sec:spec}), the chronological trend in the occurrence of emission at lower frequency favours the possibility that this feature is intrinsic to the emission mechanism. %While that is a possibility, there is a clear chronological trend in the occurrence of emission at lower frequency which favours the possibility that this feature is intrinsic to the emission mechanism. A variable spectral index between components would have resulted in emission components occurring at different frequencies without any time ordering. 
Furthermore, when the second burst is corrected for dispersion using the structure-maximising DM, the trailing component shows evidence for being under-dedispersed. Similar behaviour has been seen in a small sample of radio bursts of multiple repeating FRBs (e.g., \citealt{Caleb20, Pleunis21, Marthi22}). These observables further add credence to the suggestion that these bursts share the same emission mechanism with the FRB-like burst events seen from SGR J1935+2154~\citep{Kirsten21, Giri23, Zhu23}.

We compare the isotropic-equivalent spectral luminosity of the brightest component in the two bursts detected by Murriyang and MeerKAT from \xmmsrc\ with other short-duration radio transients in Figure~\ref{fig:phase_space}. %We estimate that the brightest component in the burst discovered by Murriyang is two orders of magnitude brighter than the one seen in MeerKAT. 
The brightest component in the MeerKAT burst is comparable in luminosity to the brighter pulsars. This is also consistent with the typical radio fluxes observed from radio magnetars. %radio emission from so-called magnetars, believed to have a large magnetic field compared to the canonical pulsar population. 
On the other hand, the Murriyang burst is two orders of magnitude brighter than the MeerKAT burst, similar to the luminosity jumps seen in the bright bursts from SGR~J1935+2154 \citep{Kirsten21}, the neutron star that emitted two FRB-like bursts and comparable to the bright radio burst from another magnetar 1E 1547.0$-$5408~\citep{israel2021}. Given the luminosity of the Murriyang burst and the observational similarities to repeating FRBs, we speculate that these bursts may share the same emission mechanism to sporadic bright bursts observed from SGR J1935+2154~\citep{Kirsten21}. 
If we assume that the magnetic field powers these bursts~\citep{Lyubarsky20, Yuan20}, we predict that the same mechanism operates over multiple orders of magnitude in magnetic field strength. Assuming that the radio luminosity is some fraction of the total magnetic energy of the star ($\sim\, k\int \frac{B^{2}(r)}{2\mu_{0}}~d^{3}r$ where $k\ll 1$), the expected luminosity of the burst can be at least 4-5 orders of magnitude smaller than the brighter bursts seen from SGR J1935+2154 if the magnetic field strength for \xmmsrc\ is a factor of 100 smaller. This is consistent with the estimated luminosities for the bursts presented in this work. %We do note however that the bursts may also follow a luminosity distribution similar to the distribution seen in other FRBs~\citep{} that depend on various physical paramters (e.g. magnetic field strength, plasma density). Hence, a single burst may not directly reflect the true physical conditions of the emitting region. 

%While no periodic signal was detected in the Murriyang observations, we estimate the spin period ($P$) of \xmmsrc\ using the empirical relation $\tau(\text{ms})\sim0.6\times P(\text{s})$, where $\tau$ is the typical microstructure timescale \citep{Kramer24}. Given our observed $\tau=1.1$\,ms, we infer $P\sim2$\,s. Such a spin period would be consistent with XDINSs and magnetars.
We estimate the radio burst rate of \xmmsrc\ to be $0.05^{+0.13}_{-0.04}\,\text{hr}^{-1}$ above the Murriyang sensitivity limit of $\sim0.3$\,Jy. The uncertainties are computed at 95\% confidence level assuming Poisson statistics. This burst rate is higher than those estimated from the monitoring campaign of magnetars with a sensitivity limit of $\sim7.5$\,Jy at 408\,MHz ($<226\,\text{--}\,785\,\text{yr}^{-1}$; \citealt{Geminardi25}).
\section{Summary}\label{sec:sum}
In this paper, we report the detection of a bright and wideband radio burst from \xmmsrc\ using the Murriyang UWL receiver. The burst is two orders of magnitude brighter than the original burst detected by MeerKAT 14 months prior. The burst shows a fast rising, slowly decaying morphology and emission across the full UWL bandwidth with a spectral index of $\alpha=-2.18\pm0.16$. We find microstructure in the burst on a typical timescale of 1.1\,ms with no evidence of periodicity. Polarimetric analysis shows the burst is 54\% linearly polarised and 22\% circularly polarised. We identify an orthogonal jump in the PPAs, consistent with the emission arising from the neutron star's magnetosphere. The peak luminosity is brighter than most pulsars and comparable to some of the bright bursts of SGR 1935+2154. If this emission is magnetically powered, it suggests that the underlying emission mechanism could operate over a wide range of magnetic field strengths and radio luminosities. Our discovery of two bursts from \xmmsrc\ -- detected by Murriyang and MeerKAT, respectively -- demonstrates that radio-quiet neutron stars such as XDINSs and CCOs may emit rare, sporadic radio emission and highlights the need for regular monitoring of these objects.

\section*{Acknowledgements}
The Parkes radio telescope (Murriyang) is part of the Australia
Telescope National Facility that is funded by the Australian Government for operation as a National Facility managed by CSIRO. We acknowledge the Wiradjuri people as the traditional owners of the Observatory site.
JT thanks Shi Dai for support and guidance with the flux calibration of the UWL data. KMR acknowledges support from a UKRI-STFC grant (SKA-NIPS, no. ST/Z510439/1). JT and BWS acknowledge funding from an STFC Consolidated grant. MC and KS acknowledge support of an Australian Research Council Discovery Early Career Research Award (project number DE220100819) funded by the Australian Government.
IPM further acknowledges funding from an NWO Rubicon Fellowship, project number 019.221EN.019.
This research has made use of NASA’s Astrophysics Data System Bibliographic Services.
%%%%%%%%%%%%%%%%%%%%%%%%%%%%%%%%%%%%%%%%%%%%%%%%%%
\section*{Data Availability}
The raw UWL data will be available to download via the CSIRO Data Acess Portal (\url{https://data.csiro.au/}) following an 18-month proprietary period. Other data products will be shared on reasonable request to the corresponding author.
%%%%%%%%%%%%%%%%%%%% REFERENCES %%%%%%%%%%%%%%%%%%

% The best way to enter references is to use BibTeX:

\bibliographystyle{mnras}
\bibliography{references,bibfile} 

% Alternatively you could enter them by hand, like this:
% This method is tedious and prone to error if you have lots of references
%\begin{thebibliography}{99}
%\bibitem[\protect\citeauthoryear{Author}{2012}]{Author2012}
%Author A.~N., 2013, Journal of Improbable Astronomy, 1, 1
%\bibitem[\protect\citeauthoryear{Others}{2013}]{Others2013}
%Others S., 2012, Journal of Interesting Stuff, 17, 198
%\end{thebibliography}

%%%%%%%%%%%%%%%%%%%%%%%%%%%%%%%%%%%%%%%%%%%%%%%%%%

%%%%%%%%%%%%%%%%% APPENDICES %%%%%%%%%%%%%%%%%%%%%

\appendix
\section{Timeline of the Murriyang observations}\label{append:timeline}
\begin{figure*}
\centering
\includegraphics[width=0.6\textwidth]{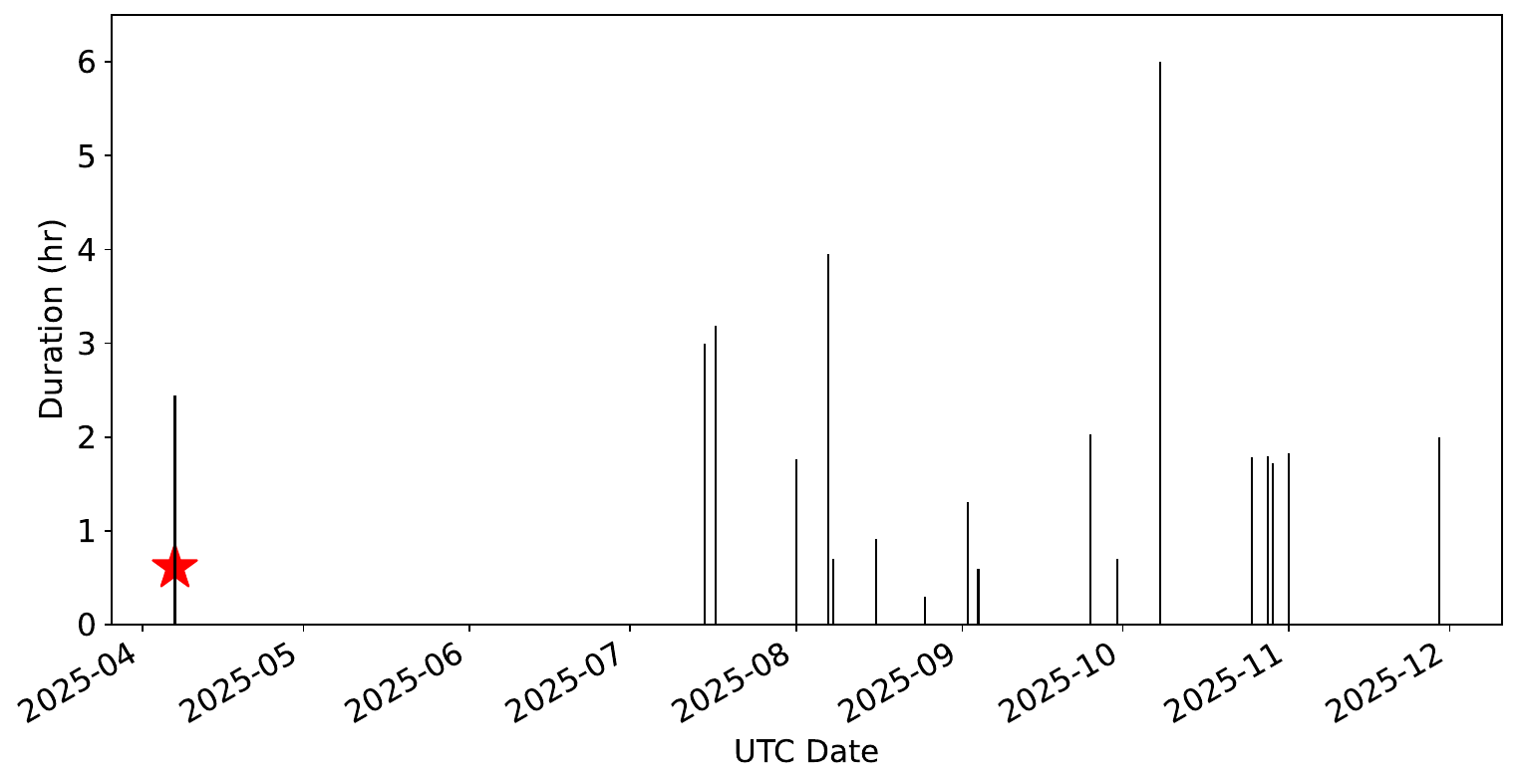}
\caption{A timeline of targeted Murriyang observations of 2XMM J104608.7$-$594306. Each vertical line shows the time of an observation, and the height of the line shows the observation duration. The burst detection is marked by a red star.}
\label{fig:obs}
\end{figure*}

Figure~\ref{fig:obs} shows the timeline of the Murriyang observations of \xmmsrc, spanning from April 2025 to November 2025. The detection of the bright radio burst reported in this paper is marked by a red star.
\section{Comparison of the Murriyang and MeerKAT bursts}\label{append:compare}
\begin{figure}
	\centering
	\includegraphics[width=0.5\textwidth]{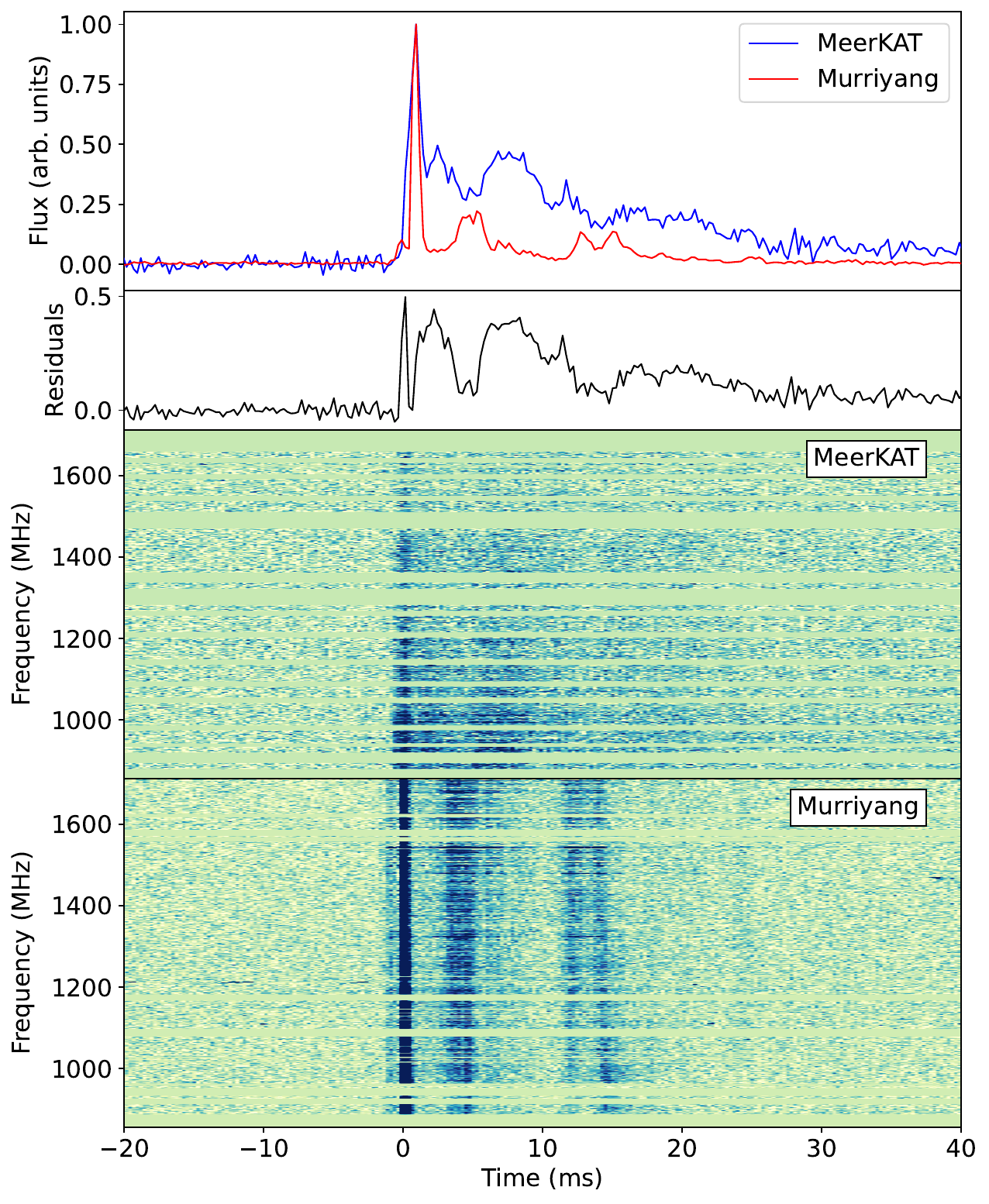}
	\caption{Burst comparison from \xmmsrc\ between the MeerKAT and Murriyang bursts over the same bandwidth and at the same spectro-temporal resolution. The top panel shows the pulse profiles averaged over 856--1712\,MHz, normalised to the same main peak amplitude (arbitrary units). 
    The second panel displays the residuals after subtracting the pulses.
    The third panel shows the MeerKAT burst resampled to the Murriyang spectro-temporal resolution, and the bottom panel shows the Murriyang burst at the MeerKAT bandwidth.
	}
	\label{fig:compare}
\end{figure}

In Figure~\ref{fig:compare} we show the pulse profiles of the two bursts detected by Murriyang and MeerKAT, respectively, with their peaks aligned in time. As the Murriyang observation has a much wider bandwidth (704--4032\,MHz) than MeerKAT (856--1712\,MHz), we chose to show the pulse averaged over the same frequency range between 856--1712\,MHz. We also downsampled the MeerKAT data to the same time resolution as Murriyang, i.e. $256\,\mu$s. As the Murriyang burst is much more luminous than the MeerKAT burst, we normalised both bursts by their respective peaks. The second panel shows the difference between the two normalised pulse profiles.
% \begin{figure}
% 	\centering
% 	\includegraphics[width=0.5\textwidth]{bursts_compare.pdf}
% 	\caption{Pulse profiles of the Parkes and MeerKAT detected bursts from \xmmsrc\ averaged over 856--1712\,MHz and with a time resolution of $256\,\mu$s. We normalise the pulses by their respective peaks. The bottom panel displays the residuals after subtracting the two pulses in arbitrary units.
% 	}
% 	\label{fig:compare}
% \end{figure}

\bsp	% typesetting comment
\label{lastpage}
\end{document}